\documentclass[twocolumn,bibyear]{aa} 
\usepackage{graphicx}
\usepackage{txfonts}
\usepackage{epsfig}
\usepackage{natbib}
\usepackage{graphicx}
\usepackage{multirow}
\usepackage{lscape}
\usepackage{mathrsfs,amssymb}
\usepackage{amsmath}
\usepackage{lineno}
\newcommand       \Angstrom     {\,{\rm \AA}}

\newcommand       \cm           {\,{\rm cm}}

\newcommand       \erg          {\,{\rm erg}}
\newcommand       \eV           {\,{\rm eV}}

\newcommand       \s            {\,{\rm s}}

\newcommand       \Myr      {\,{\rm Myr}}

\newcommand       \simgt        {\gtrsim}

\newcommand       \mum          {\,{\rm \mu m}}
\newcommand       \ppm          {\,{\rm ppm}}

\newcommand       \msun         {\,{{\rm M}_\odot}}
\newcommand       \Msun         {\,{{\rm M}_\odot}}

\newcommand       \simali       {\sim\,}
\newcommand       \magni        {\,{\rm mag}}

\newcommand       \NC          {N_{\rm C}}
\newcommand       \NH          {N_{\rm H}}
\newcommand       \NPAH          {N_{\rm PAH}}

\newcommand       \CTOHPAHa  {\left[{\rm C/H}\right]_{\rm\scriptsize PAH}}
\newcommand       \CTOHPAHb  {\left[{\rm\frac{C}{H}}\right]_{\rm\scriptsize PAH}}
\newcommand       \CTOHgraa  {\left[{\rm C/H}\right]_{\rm gra}}

\newcommand       \etapah  {\eta_{\rm\scriptsize PAH}}
\newcommand       \etaPAH  {\eta_{\rm\scriptsize PAH}}
\newcommand       \Fobs  {F_\lambda^{\rm obs}}
\newcommand       \Fintr  {F_\lambda^{\rm intr}}
\newcommand       \Fmod  {F_\lambda^{\rm mod}}
\newcommand       \taulambda  {\tau_\lambda}

\newcommand       \Ndata  {N_{\rm data}}
\newcommand       \sigobs  {\sigma_{{\rm obs},\lambda}}
\newcommand       \sigobsp  {\sigma^{\prime}_{{\rm obs},\lambda}}
\newcommand       \omegap  {\omega^{\prime}}
\newcommand       \pitopi  {\pi^{\ast}\leftarrow\pi}
\newcommand       \widthWL  {\Delta\lambda}
\newcommand       \widthWV  {\gamma}

\begin{document}

\title{Polycyclic aromatic hydrocarbon
        and the ultraviolet extinction bump
        at the cosmic dawn}
\titlerunning{Ultraviolet extinction bump at the cosmic dawn}
   \author{Qi~Lin\inst{1,2},
        X.J.~Yang\inst{1,2},
        Aigen Li\inst{2}, \and
        Joris~Witstok\inst{3,4}
          }
\authorrunning{Q. Lin et al.}
\institute{Hunan Key Laboratory for Stellar and Interstellar Physics and School of Physics and Optoelectronics, Xiangtan University, Hunan 411105, China\\
\email{xjyang@xtu.edu.cn}
\and
Department of Physics and Astronomy,
             University of Missouri,
             Columbia, MO 65211, USA \\
\email{lia@missouri.edu}
\and
Cosmic Dawn Center (DAWN), Copenhagen, Denmark
\and
Niels Bohr Institute, University of Copenhagen, Jagtvej 128, DK-2200, Copenhagen, Denmark
             }
\date{Received XX XX, XXXX; accepted XX XX, XXXX}
\abstract
   {
First detected in 1965, the mysterious ultraviolet 
(UV) extinction bump at 2175$\Angstrom$
is the most prominent spectroscopic feature
superimposed on the interstellar extinction curve.
Its carrier has remained unidentified
over the six decades
since its first detection,
although many candidate materials
  have been proposed.}
  {Widely seen in the interstellar medium of
the Milky Way as well as several nearby galaxies,
this bump was recently also detected by
the \textit{James Webb} Space Telescope (JWST)
at the cosmic dawn in JADES-GS-z6-0,
a distant galaxy at redshift $z\approx6.71$,
corresponding to a cosmic age of just 800
million years after the big bang.
Differing from that of the known Galactic
and extragalactic interstellar sightlines, 
which 
always peak at $\simali$2175$\Angstrom$, 
the bump seen at $z\approx6.71$
peaks at an appreciably longer wavelength of 
$\simali$2263$\Angstrom$
and is the narrowest among all known 
Galactic and extragalactic extinction bumps.}
   {Here we show that the combined electronic
absorption spectra quantum chemically
computed for a number of polycyclic aromatic
hydrocarbon (PAH) molecules closely reproduce
the bump detected by JWST in JADES-GS-z6-0.}
   {This suggests that PAH molecules had already
been pervasive in the Universe
at an epoch when asymptotic giant branch
stars had not yet evolved to make dust.} 
   {}
\maketitle
\section{Introduction\label{sec:intro}}
First detected nearly six decades ago
by Stecher (1965), the ultraviolet (UV) extinction
bump at 2175$\Angstrom$ is the most prominent
spectral feature on the interstellar extinction curve.
Such a UV bump is widely seen in the Milky Way
and nearby galaxies, including the Large Magellanic
Cloud, several regions in the Small Magellanic Cloud, and M31 (see Henning \& Schnaiter 1998, Draine 2003).
Despite nearly 60 years of extensive observational,
theoretical, and experimental studies, the exact carrier
of the 2175$\Angstrom$ extinction bump
remains unidentified.
Although a variety of candidate materials
have been proposed (see Wang et al.\ 2023
and references therein),
consensus has yet to be reached.

Utilizing the Near Infrared Spectrograph (NIRSpec)
on board JWST, Witstok et al.\ (2023) detect
an extinction bump at the cosmic dawn
in JADES-GS+53.15138-27.81917
(also known as JADES-GS-z6-0), 
a distant galaxy at redshift $z\approx6.71$.
This indicates that the bump carriers
had already been pervasive in the Universe
at a cosmic age of just 800 million years (Myr)
after the big bang.
At such an early epoch of cosmic time
there was not enough time for low- to
intermediate-mass stars ($\simali$0.5--8$\msun$)
to evolve sufficiently to the dust-producing
asymptotic giant branch (AGB) phase.
The formation of the bump carriers
during such a short (approximately 400$\Myr$) phase
of stellar evolution provides valuable insight
into the nature of the bump carriers, as well as
the stellar evolution and dust condensation
scenarios in the earliest epochs of the Universe.
The 2175$\Angstrom$ bump has also been
detected at the cosmic noon in galaxies
at $z$\,$\approx$1--2 (e.g., see Shivaei et al.\ 2022).

Immediately after its first detection (Stecher 1965),
small graphite grains were proposed to be responsible
for the 2175$\Angstrom$ extinction curve
(Stecher \& Donn 1965).
However, as Li et al.\ (2024) demonstrate,
graphitic grains fail to explain the UV bump
detected in JADES-GS-z6-0,
as the extinction bump arising from those grains
is too broad and peaks at wavelengths that are
too long to agree with what is seen
in JADES-GS-z6-0.\footnote{%
   Strictly speaking, small, nano-sized graphite
grains with a very definite graphite structure
do not exist; instead, carbon nanoparticles may contain
very small crystalline, turbostratic graphitic
basic structural units (see J\"ager et al.\ 2008).
The UV absorption spectra measured for such
nano-sized carbon grains do exhibit a prominent
bump that closely resembles the 2175$\Angstrom$
extinction bump (Schnaiter et al.\ 1998).
  }
In this work, we show that the quantum chemically
computed electronic absorption spectra of a large
number of polycyclic aromatic hydrocarbon (PAH)
molecules closely reproduce the UV extinction bump
seen in  JADES-GS-z6-0 at $z\approx6.71$.
This paper is organized as follows.
In \S\ref{sec:methods}, we briefly describe
the computational methods
and target PAH molecules.
The peak wavelengths and widths of
the UV bumps calculated for the target molecules
are also presented in \S\ref{sec:methods}.
We describe in \S\ref{sec:model}
how the computed electronic absorption spectra
are utilized to model the UV extinction bump
seen in JADES-GS-z6-0.
The results are presented in \S\ref{sec:results}
and discussed in \S\ref{sec:discussion}.
Our major conclusion is summarized
in \S\ref{sec:summary}.
%

\begin{figure*}
        \centering
\includegraphics[width=0.98\textwidth]{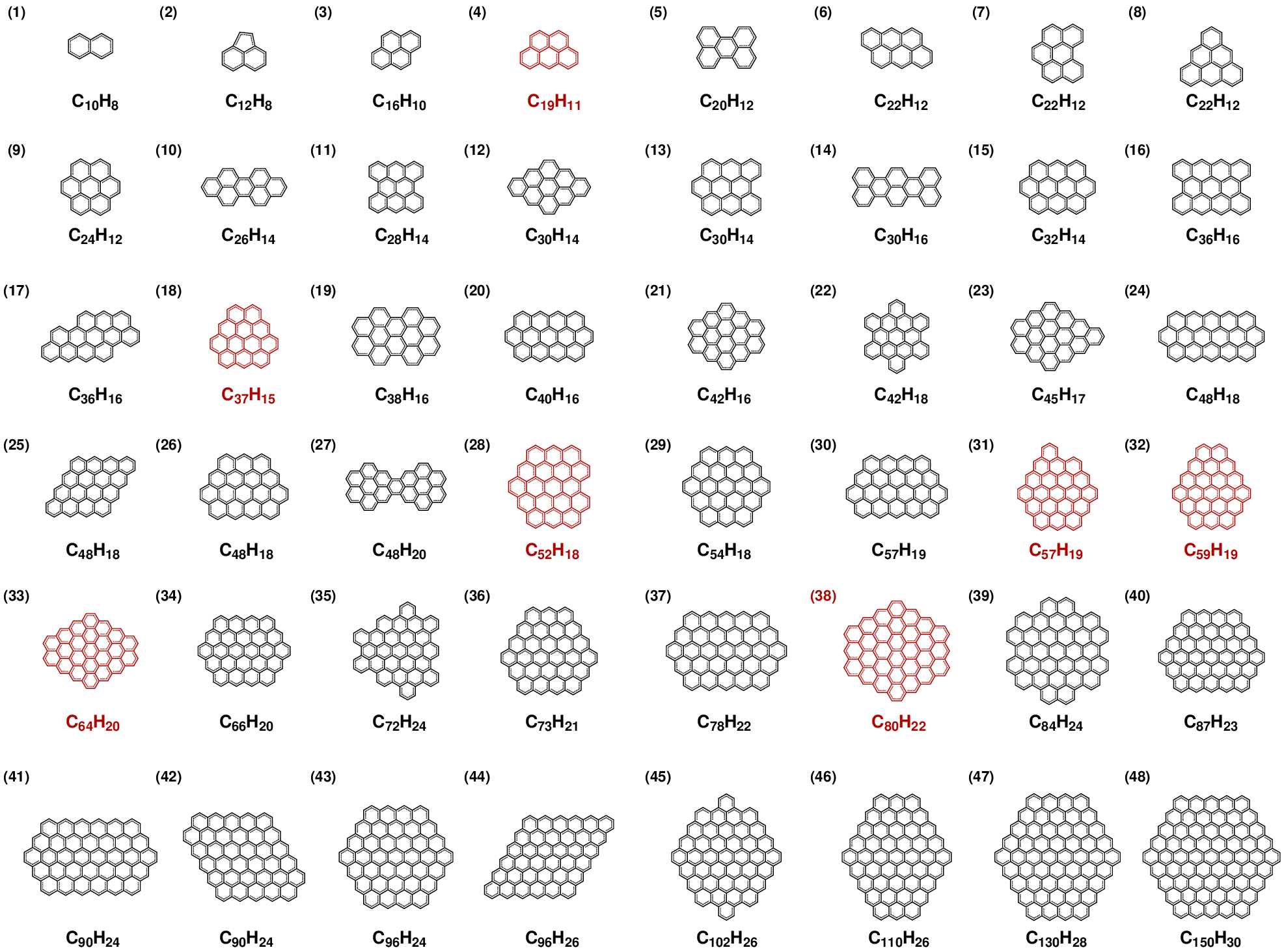}
\caption{\label{fig:48molecules}
           Target PAH molecules for TD-DFT computations
           of their electronic transitions.
           }
\end{figure*}

\section{Computational methods and
        target PAH molecules}\label{sec:methods}
We considered 48 PAH molecules spanning a wide range
of sizes from 10 carbon (C) atoms
($\NC=10$; naphthalene, C$_{10}$H$_{8}$)
to 150 C atoms (circumcircumcircumcoronene,
C$_{150}$H$_{30}$).
The sizes of these molecules are roughly evenly
distributed, except for the largest ones with $\NC>100$.
As illustrated in Fig.~\ref{fig:48molecules},
most of these molecules have
a compact, pericondensed structure.
In the interstellar medium (ISM),
pericondensed PAHs are structurally stable
in terms of thermal and chemical properties
among the various isomers.

It has been argued that only PAHs
with $\NC\simgt20$ can survive in the ISM
(see Tielens 2008).
However, Stockett et al.\ (2023) find that nitrile-substituted naphthalene
(i.e.,  cyanonaphthalene, C$_{10}$H$_7$CN)
can be efficiently stabilized
with the aid of the so-called recurrent fluorescence
(also known as Poincar\'e fluorescence,
see L\'eger et al.\ 1988),
a radiative relaxation channel
in which optical photons are emitted
from thermally populated electronically excited states.
Iida et al.\ (2022) also find that the recurrent
fluorescence could efficiently stabilize
small cationic carbon clusters
with as few as nine carbon atoms.
Indeed, a number of small, specific PAH
molecules---indene
(C$_9$H$_8$; Burkhardt et al.\ 2021,
Cernicharo et al.\ 2021),
cyanoindene (C$_9$H$_7$CN; Sita et al.\ 2022),
cyanonapthalene (McGuire et al.\ 2021),
and cyanopyrene
(C$_{16}$H$_9$CN; Wenzel et al.\ 2024a,b)---as
well as benzonitrile (C$_6$H$_5$CN; McGuire et al.\ 2018)
have been discovered in the Taurus Molecular Cloud
on the basis of their rotational transitions
at radio frequencies.
In addition, Gredel et al.\ (2011) searched for
the electronic absorption fingerprints
in the UV/visible 
wavelengths 
of small molecules including anthracene,
phenanthrene, and pyrene, and derived
upper limits on their abundances.
%
Therefore, we included four molecules
with $\NC<20$ in our sample.
On the other hand, we did not consider
molecules with more than 150 C atoms.
First, these molecules are computationally
expensive. Also, as will be shown later,
larger molecules (with $\NC>60$)
tend to shift their $\pi^{*}\gets\pi$
electronic transitions to longer wavelengths.
As the peak wavelengths of the $\pi^{*}\gets\pi$
transitions of molecules with $\NC>100$
substantially deviate from the UV extinction
bump seen in JADES-GS-z6-0,
it is not necessary to consider molecules
with $\NC>150$.

Unfortunately, there is a lack of experimental measurements
of the UV/visible absorption spectra of gas-phase,
intermediate-sized PAH molecules
(with $\NC$\,$\simali$20--60).
Even for small molecules (with $\NC<20$),
gas-phase measurements
are also very limited (e.g., see Gredel et al.\ 2011).
Therefore, we relied on quantum
chemical computations.
However, the substantial quantity of electrons
present in the PAH molecules under examination
(reaching 930 in the largest molecule,
circumcircumcircumcoronene)
currently hinders any ab initio investigation
that directly solves the many-electron
Schr\"odinger equation, due to excessively
high computational expenses.
In this study, we resorted to density functional theory
(DFT; Jones \& Gunnarsson 1989)
and its time-dependent version
(TD-DFT; Marques \& Gross 2004)
to examine the ground-state and
excited-state characteristics of PAHs.\footnote{%
  The accuracy of the TD-DFT method
  has been verified by the experimental,
  gas-phase spectrum of anthracene
  (C$_{14}$H$_{10}$; see Malloci et al.\ 2004).
  However, it is not clear how accurate
  this method is for large molecules.
  We urgently need gas-phase experimental
  UV/visible spectra of PAH molecules of
  different species and sizes.
  }

We first performed structure optimization
utilizing the GAUSSIAN 16 software (Frisch et al.\ 2016).
We used the hybrid density functional model (B3LYP)
at the 6-31+G(d) level.
This method provides sufficient calculation accuracy
with operable computer time for large molecules.
We simultaneously calculated the vibrational modes
of each species to confirm that the optimized structures
are at the minima of their potential energy surfaces.

We then used the real-space, real-time version of
TD-DFT, as implemented in the code OCTOPUS
(Marques et al.\ 2003), to compute the UV/visible
electronic spectra of the optimized structures.
We employed a real-space grid comprising overlapping
spheres with a radius of 3$\Angstrom$ centered
around each atom. All quantities were discretized
in a uniform grid with a spacing of 0.3$\Angstrom$.
The time step for the time evolution was
0.002\,$\hbar\eV^{-1}$ ($\hbar$ is the reduced
Planck constant), which ensures stability
in the time-dependent propagation
of the Kohn-Sham wave functions.
Furthermore, the total propagation time
was at least 20\,$\hbar\eV^{-1}$.

\begin{figure*}
        \centering
\includegraphics[width=0.98\textwidth, angle=0]{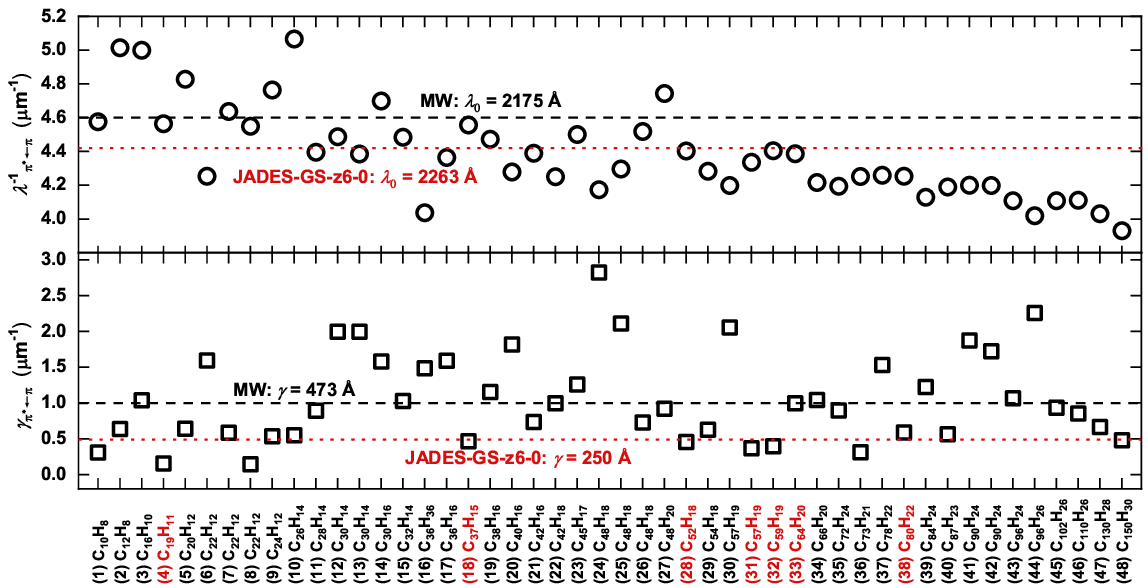}
\caption{
              \label{fig:lambdawidth}
              Peak wavelengths and widths of
               $\pi^{\ast}$\,$\leftarrow$\,$\pi$ transitions
              of individual PAH species.
              Also shown are those of the Galactic average
              (horizontal, dashed black lines)
              and JADES-GS-z6-0 observed by JWST/NIRSpec
              (horizontal, dashed red lines).
         }
\end{figure*}

For each molecule, we calculated the electronic
absorption cross section, $C_{\mathrm{abs}}(\lambda)$.
We then fitted the calculated $C_{\mathrm{abs}}(\lambda)$
by a Drude function combined with a Fano profile
(see Lin et al.\ 2023).
The peak position ($\lambda _{\pi^{*}\gets \pi }^{-1} $)
and width ($\gamma_{\pi^{*}\gets \pi }$) of the Drude
function characterize the electronic $\pi^{*}\gets \pi$
transition, which 
resembles the UV extinction bump 
seen in the ISM. In Fig.~\ref{fig:lambdawidth} we
show $\lambda _{\pi^{*}\gets \pi }^{-1} $
and $\gamma_{\pi^{*}\gets \pi }$
for each molecule.

In general, as shown in Fig.~\ref{fig:lambdawidth}, 
the absorption bumps for the $\pi^{*}\gets\pi$
transitions for larger PAH molecules
(with $\NC>60$) tend 
to peak at somewhat longer wavelengths. 
In contrast, their bump widths are
randomly distributed around
$\gamma=1.0\mum^{-1}$,
and the variation of the bump width with
PAH size shows no systematic tendency.
As will be elaborated on in \S\ref{sec:model},
the extinction bump seen in JADES-GS-z6-0
is very narrow (at only half of the mean
width of the Galactic bump) and ``red''
(which peaks at $\simali$2263$\Angstrom$,
a wavelength appreciably longer than
the Galactic nominal wavelength of
$\simali$2175$\Angstrom$).
In the following, we will confront
the computed electronic absorption
spectra of PAH mixtures with
these observational characteristics
to assess the viability of PAHs
as a potential explanation
for the extinction bump
detected in JADES-GS-z6-0.

\section{Extinction bump modeling}\label{sec:model}
JADES-GS-z6 is a low-metallicity,
normal star-forming galaxy
with a subsolar metallicity
of $Z$\,$\approx$\,0.2--0.3\,$Z_{\odot}$
(Witstok et al.\ 2023).
The Balmer decrement demonstrates that
this galaxy suffers significant dust obscuration.
The rest-frame UV spectrum obtained by
JWST/NIRSpec exhibits a pronounced dip at wavelength
$\lambda_0$\,$\simali$2263$^{+20}_{-24}\Angstrom$,
with a width (measured in wavelength) of
$\Delta\lambda$\,$\simali$250$\Angstrom$
(which corresponds to
$\widthWV$\,$\simali$0.49$\mum^{-1}$
in inverse wavelength).
This dip resembles the interstellar UV extinction
bump but shifting the nominal peak wavelength
of 2175$\Angstrom$ to a somewhat longer
wavelength. Its width is considerably narrower
than the mean width of
$\widthWL\simali$473$\Angstrom$
(or $\widthWV\simali$1$\mum^{-1}$)
of the Galactic ISM (Valencic et al.\ 2004).
In the following, we shall still refer to this dip
as the 2175$\Angstrom$ bump,
despite the fact that it actually peaks at 2263$\Angstrom$.

If we let $\Fintr$ be the intrinsic, extinction-free
flux emitted by JADES-GS-z6 at wavelength $\lambda$
that would be ``detected'' by JWST,
and $\Fmod$ be the dust-attenuated ``model'' flux
expected to be observed by JWST/NIRSpec, then
$\Fmod$ and $\Fintr$ are related through
\begin{equation}\label{eq:FobsFintr}
        \Fmod = \Fintr \exp\left(-\taulambda\right)~~,
\end{equation}
where $\taulambda$ is the optical depth
at wavelength $\lambda$
arising from the dust along the line of sight
toward the region in JADES-GS-z6 observed
by JWST. For a mixture of $\NPAH$ different
PAH species, we derive the optical depth from
\begin{equation}\label{eq:tau1}
        \tau_\lambda = \NH\times\CTOHPAHb\times\sum_{j=1}^{\NPAH}
        \omega_j\,\left[\frac{C_{{\rm abs},j}(\lambda)}{N_{{\rm C},j}}\right] ~~,
\end{equation}
where $\NH$ is the hydrogen (H) column density
for the interstellar sightline in JADES-GS-z6
observed by JWST, $\CTOHgraa$ is the total
amount of C relative to H locked up in $\NPAH$
PAH species, $C_{{\rm abs},j}(\lambda)$ is the absorption
cross section computed for the $j$-th PAH species,
$N_{{\rm C},j}$ is the number of C atoms contained
in the $j$-th PAH species, and $\omega_j$ is
the fractional weight of the $j$-th PAH species.

To best reproduce $\Fobs$, the flux observed by JWST,
we need to minimize
\begin{equation}\label{eq:chi2}
        \chi^2 = \sum_{i=1}^{\Ndata}
        {\omegap}^2\,\left\{\frac{\ln\Fobs-\ln\Fmod}
        {\sigobsp}\right\}^2 ~~,
\end{equation}
where $\Ndata$ is the number of data points
in the wavelength range of 1993.1--2488.7$\Angstrom$
where the UV extinction bump peaks,
$\omegap$ is the weight we introduced
to enhance the fit to the bump,
and $\sigobsp$ is the reduced observational
uncertainty:
\begin{equation}\label{eq:sigmaobsp}
        \sigobsp \equiv \sigobs/\Fobs ~~,
\end{equation}
where $\sigobs$ is the observational uncertainty.
%
Over the wavelength range of
1993.1--2488.7$\Angstrom$,
the JWST/NIRSpec spectrum has 33 data points.
%
%
If we consider all 48 PAH molecules in fitting
$\Fobs$, we would have more model parameters
than the number of data points.
To avoid this, we preselected 17 (of 48) species
for which $\gamma<1.0\mum^{-1}$
and $\lambda_{\pitopi}^{-1}<4.6\mum^{-1}$
(see  Fig.~\ref{fig:lambdawidth}).
In view of the fact that the bump seen
in JADES-GS-z6-0 is very narrow
($\gamma\approx0.49\mum^{-1}$)
and peaks at an appreciably longer wavelength
($\simali$4.4$\mum^{-1}$) than the Galactic
UV bump, we made such a preselection to
ensure that the computed UV bumps of these
molecules were not too broad and their peak
wavelengths were not too short.
%
We adopted the following weight:
\begin{equation}\label{eq:weight}
        \omegap = \left\{\begin{array}{lr}
                0, & |\lambda - \lambda_0| > 2\Delta\lambda,\\
                1, & \Delta\lambda < |\lambda - \lambda_0|
                < 2\Delta\lambda,\\
                2, & \Delta\lambda/2 < |\lambda - \lambda_0|
                < \Delta\lambda,\\
                4, & |\lambda - \lambda_0| < \Delta\lambda/2,\\
        \end{array}\right.
\end{equation}
where $\lambda_0$ and $\Delta\lambda$ are
the peak wavelength and full width half maximum
(FWHM) of the ``2175$\Angstrom$'' UV extinction bump.
While rather arbitrary, we designed such a weight
to ``force'' the model spectrum to fit the bump
observed in JADES-GS-z6-0,
with increasing emphasis put on the data in
the wavelength range closer to the bump peak.

By defining $\etaPAH\equiv\NH\times\CTOHPAHa$,
the best fit is achieved under the condition of
\begin{equation}\label{eq:chi2pah}
        \partial{\chi^2}/\partial{\etapah} = 0 ~~,
\end{equation}
and from eq.\,\ref{eq:chi2pah} we derive
\begin{eqnarray}\label{eq:etapah}
        \etapah & = & \sum_{i=1}^{\Ndata}
        \left\{
        \frac{
                \left(\ln\Fintr-\ln\Fobs\right)
                \times\langle C_{\rm abs}/\NC\rangle}
        {{\sigobsp}^2}
        \right\} \\
        & \times & \left\{
        \sum_{i=1}^{\Ndata}
        \left[\frac{\langle C_{\rm abs}/\NC\rangle}{\sigobsp}
        \right]^2
        \right\}^{-1} ~~,
\end{eqnarray}
where $\langle C_{\rm abs}/\NC\rangle$,
the ``mean'' absorption cross section
(per C atom), is obtained from averaging
over that of all $N$ different PAH species:
\begin{equation}\label{eq:cabspah}
        \langle C_{\rm abs}/\NC\rangle
        = \sum_{j=1}^{\NPAH}
        \omega_j\,\left[\frac{C_{{\rm abs},j}(\lambda)}{N_{{\rm C},j}}\right] ~~.
\end{equation}

\section{Results}\label{sec:results}
Following Witstok et al.\ (2023),
we approximated the intrinsic spectrum
by a power-law (PL) spectrum that fits
the UV continuum observed by JWST/NIRSpec:
\begin{equation}\label{eq:continuum}
        \Fintr(\lambda) = \left(0.236\pm0.012\right)
        \times10^{-20}\times\left(\frac{\lambda}
        {1500\Angstrom}
        \right)^{-2.13}\erg\s^{-1}\cm^{-2}\Angstrom^{-1}~~.
\end{equation}
We fitted the JWST/NIRSpec spectrum of JADES-GS-z6-0
by attenuating the PL spectrum with a weighted
mixture of 17 PAH species.
In the extinction bump fitting there are
18 model parameters: $\etapah$, as well
as the fractional weight, $\omega_j$, for the $j$-th
molecule (where $j$\,=\,1, 2, ..., 17).

\begin{figure*}[h]
\centering
\includegraphics[width=16.0cm]{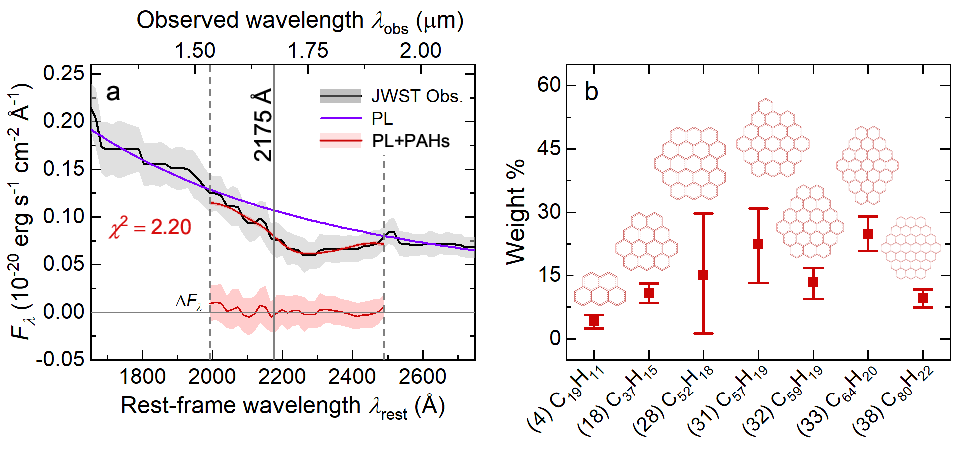}
\caption{
         \label{fig:bestfit1}
         Left panel (a): Fitting the rest-frame UV spectrum 
         of JADES-GS-z6 measured by JWST/NIRSpec
         (solid gray-shaded line) with a power-law continuum
         (solid blue line) attenuated mainly by the absorption
         of PAH molecules (solid red line).
         The bottom orange-shaded line illustrates
         the fitting residual.
         Right panel (b): The fractional weights of the
         molecules that best fit the JWST/NIRSpec
         spectrum of JADES-GS-z6.
         }
\end{figure*}

Figure~\ref{fig:bestfit1}a presents the best fit
to the JWST/NIRSpec spectrum of JADES-GS-z6-0,
as well as the derived fractional weights
for individual PAH molecules.
We see that a mixture of 17 PAH species
can closely account for the UV extinction
bump seen in JADES-GS-z6-0,
with contributions predominantly
from seven species
(in order of decreasing weights):
C$_{64}$H$_{20}$ (with a fractional weight of 24.9\%$\pm$4.1\%),
C$_{57}$H$_{19}$ (22.4\%$^{+9.1\%}_{-8.5\%}$),
C$_{52}$H$_{18}$ (15.0\%$^{+13.7\%}_{-8.5\%}$),
C$_{59}$H$_{19}$ (13.3\%$^{+3.8\%}_{-3.4\%}$),
C$_{37}$H$_{15}$ (10.8\%$\pm$2.3\%),
C$_{80}$H$_{22}$ (9.5\%$\pm$2.1\%), and
C$_{19}$H$_{11}$  (4.1\%$\pm$1.6\%).
The remaining ten species in total account
for $<$\,1\% of the fractional weight.
In Fig.~\ref{fig:bestfit1}b we show
the fractional weights of those seven
molecules that make the predominant contributions
to the UV bump. None of the other (ten) species
is shown in Fig.~\ref{fig:bestfit1}b
since their fractional weights are all
smaller than 0.1\%. 
%

Figure~\ref{fig:bestfit1} clearly shows
that the weighted sum of the electronic
absorption spectra of individual PAH molecules
fits both the peak wavelength and the width,
as well as the overall profile of the UV extinction
bump observed in the early Universe.
The best fit derives
$\etapah\equiv\NH\times\CTOHPAHa
\approx 2.34\times10^{16}\cm^{-2}$
and $\langle\NC\rangle\approx53$,
the mean size of the PAH molecules
(weighted by the relative abundances).
We took an educated estimate
for $\CTOHPAHa$, the amount of C
(relative to H) required by PAHs to
account for the observed UV bump
of JADES-GS-z6-0.
By assuming that the amount of dust
(relative to gas) is proportional to
the metallicity, $Z$, of the galaxy
and the dust size and composition
are similar to that of the Galactic diffuse ISM,
we estimated a gas-to-extinction ratio
of $\NH/E(B-V)\approx
\left(Z_\odot/Z\right)\times
\left\{\NH/E(B-V)\right\}_{\rm MW}
\approx 1.7\times10^{22}\cm^{-2}\magni^{-1}$
for JADES-GS-z6-0,
where $\left\{\NH/E(B-V)\right\}_{\rm MW}
\approx 5.8\times10^{21}\cm^{-2}\magni^{-1}$
is the mean gas-to-extinction ratio
of the Milky Way (Bohlin et al.\ 1978)
and $Z\approx0.34\pm0.05\,Z_\odot$
is the metallicity of JADES-GS-z6-0
(Witstok et al.\ 2023).
With $E(B-V)\approx0.25\pm0.07\magni$
(Witstok et al.\ 2023), we estimated
$\NH\approx4.3\times10^{21}\cm^{-2}$
and this translates to
$\CTOHPAHa\approx5.5\ppm$,
about ten times smaller than that
of the Milky Way
(i.e., $\CTOHPAHa$\,$\approx$\,40--60$\ppm$,
see Li \& Draine 2001 and Lin et al.\ 2023).\footnote{%
  This is consistent with the comparison
  in terms of $\Delta A_{2175}/\NH$,
  where $\Delta A_{2175}$ is the excess
  extinction at 2175$\Angstrom$:
with $\Delta A_{2175}\approx0.43\pm0.07\magni$
(Witstok et al.\ 2023)
and $\NH\approx4.3\times10^{21}\cm^{-2}$
as estimated above,
we derive $\Delta A_{2175}/\NH\approx1.0
\times10^{-22}\magni\cm^2$
for  JADES-GS-z6-0,
about six times lower
than the Galactic value of
$\Delta A_{2175}/\NH\approx5.8
\times10^{-22}\magni\cm^2$ (Draine 1994).
}

\begin{figure*}[h]
\centering
\includegraphics[width=16.0cm]{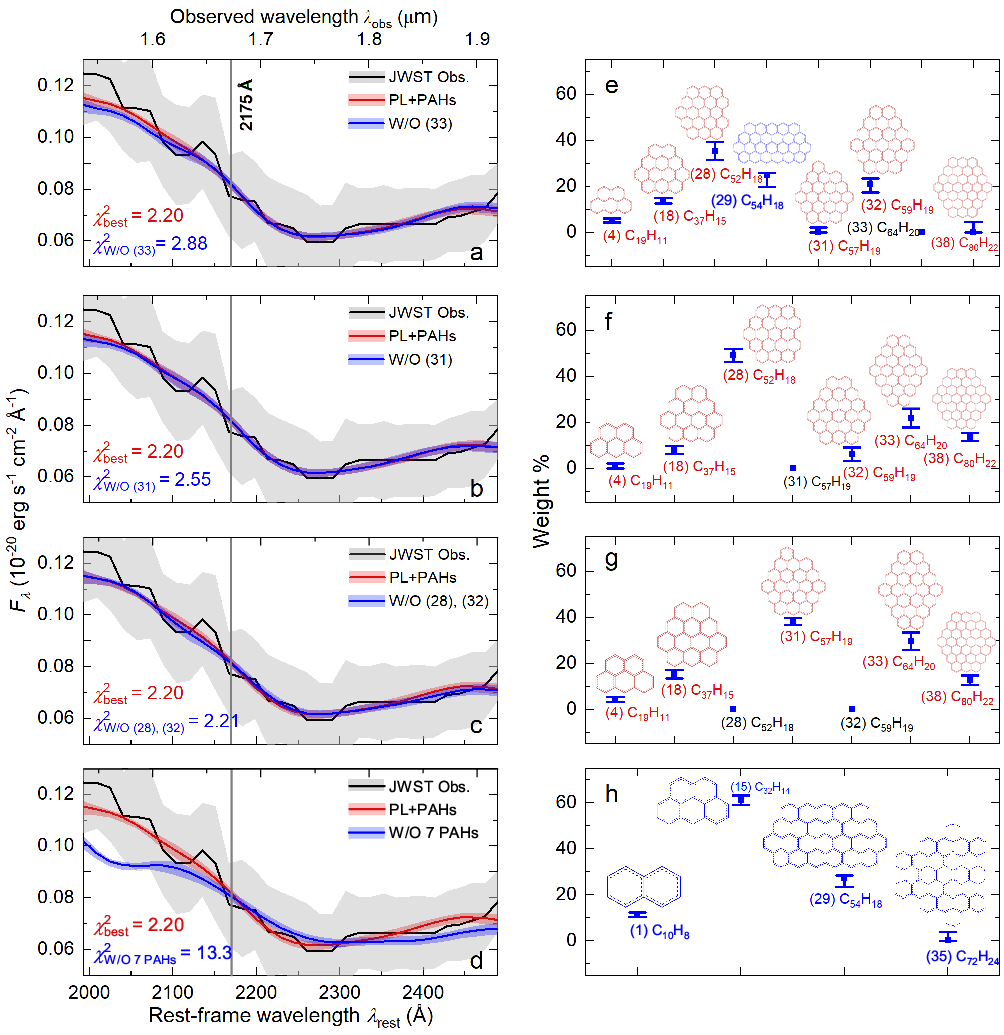}
\caption{
         \label{fig:uniqueness}
         Upper panel (a, e): Same as Fig.~\ref{fig:bestfit1}
         but with C$_{64}$H$_{20}$ excluded
         and C$_{52}$H$_{18}$ added.
         Middle panel (b, f): Same as Fig.~\ref{fig:bestfit1}
         but with C$_{57}$H$_{19}$ excluded.
         Middle panel (c, g): Same as Fig.~\ref{fig:bestfit1}
         but with C$_{52}$H$_{18}$ and C$_{59}$H$_{19}$
         excluded. Bottom panel (d, h): Same as
         Fig.~\ref{fig:bestfit1} but with all those seven
         best-fitting molecules excluded.
         }
\end{figure*}

\section{Discussion}\label{sec:discussion}
To examine the uniqueness of the fit
(especially the uniqueness of the role
of individual PAH molecules),
we explored whether satisfactory fits to the observed spectrum
can still be achieved when one or more of
the best-fitting seven molecules
(see Fig.~\ref{fig:bestfit1}) is excluded.
For this purpose, we repeated the UV bump fitting
by minimizing $\chi^2$
(see eqs.\,\ref{eq:chi2},\,\ref{eq:etapah})
and excluding one or more molecules
from the fitting. We first excluded C$_{64}$H$_{20}$,
the molecule with the highest fractional weight
as derived above.
As shown in Fig.~\ref{fig:uniqueness}a,
a close fit can still be obtained
if another molecule, that is, C$_{54}$H$_{18}$
(with a fractional weight of 24.7\%), is included.
Figure~\ref{fig:uniqueness}b shows that,
if C$_{57}$H$_{19}$ (the second most abundant
molecule among the seven best-fit molecules)
is excluded, a close fit can also be obtained
even without the inclusion of additional molecules.

We also considered the removal of C$_{52}$H$_{18}$
and C$_{59}$H$_{19}$. As the $\pitopi$ bumps of these
two molecules have similar peak wavelengths and widths,
we simultaneously excluded both C$_{52}$H$_{18}$
and C$_{59}$H$_{19}$. As shown in Fig.~\ref{fig:uniqueness}c,
a satisfactory fit to the observed UV bump
can still be achieved with the remaining five molecules.
This implies that the UV bump alone does not allow one
to uniquely pinpoint to an individual molecule.
It would be interesting to explore the infrared (IR)
emission spectra of these molecules and see
if the combination of the UV bump and the IR
emission spectra can uniquely infer the presence
of individual molecules and their abundances.
Nevertheless, as shown in Fig.~\ref{fig:uniqueness}d,
if all seven of the molecules are excluded, then no satisfactory
fits can be obtained.


As mentioned earlier in \S\ref{sec:methods},
while the vibrational bands of PAHs in the IR
are not molecule specific, their rotational transitions
occurring in radio frequencies provide characteristic
fingerprints. Indeed, a number of small PAH molecules
(i.e., indene, cyanoindene, cyanonapthalene,
and cyanopyrene), as well as benzonitrile,
have been identified in space
through their rotational lines.
Similarly, individual PAH molecules have sharp
absorption features in the UV and these electronic
transitions are characteristic of the PAH molecular
structure, so that they could potentially also
allow for the identification of specific species
(e.g., see Salama et al.\ 2011, Gredel et al.\ 2011).
However, if PAHs are present in the ISM
with a wide variety of molecules and ions,
a blend of absorption bands from
such a mixture of various species would
smooth out the sharp features from individual
PAH molecules, as demonstrated experimentally
(see Joblin et al.\ 1990, Steglich et al.\ 2010, 2012).
%
Indeed, the search for characteristic absorption
features of individual PAHs superposed on
the interstellar extinction curves was not
successful (e.g., see Clayton et al.\ 2003,
Gredel et al.\ 2011).
Nevertheless, we stress that, although the UV bump
does not allow us to uniquely identify individual molecules,
as shown in Fig.~\ref{fig:uniqueness}d,
if all seven molecules are excluded,
then no satisfactory fits can be obtained.

\begin{figure*}[h]
\centering
\includegraphics[width=16.0cm]{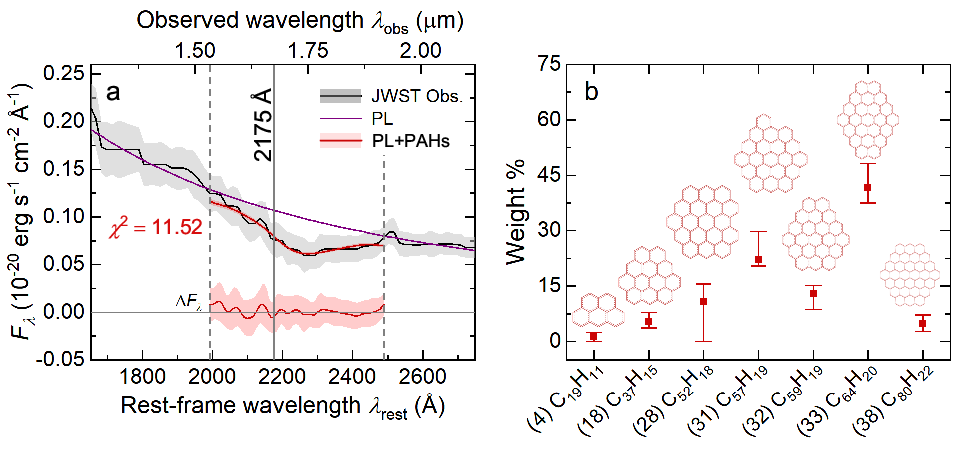}
\caption{
  \label{fig:bestfit2}
  Same as Fig.~\ref{fig:bestfit1}, but with all 48 PAH
  species included in fitting the rest-frame UV spectrum
  of JADES-GS-z6 measured by JWST/NIRSpec.
           }
\end{figure*}

We had thus far confined ourselves to
a preselection of 17 (of 48) PAH species
for which $\gamma<1.0\mum^{-1}$
and $\lambda_{\pitopi}^{-1}<4.6\mum^{-1}$.
To examine the validity of this preselection,
we also fitted the UV bump detected
in JADES-GS-z6-0 in terms of a complete
sample of all 48 PAH species.
In this case, to ensure the number of data points
would exceed the number of model parameters
($N_{\rm mod}=49$), we applied cubic spline
interpolation to expand the original 33 data
points to $\Ndata=200$.
As shown in Fig.~\ref{fig:bestfit2},
the JWST/NIRSpec spectrum of JADES-GS-z6-0
is closely fitted by a mixture of 48 PAH species,
with contributions predominantly
from exactly the same seven species
as those based on the mixture of
17 preselected molecules.
However, the derived fractional weights
of these molecules are different.
This also indicates that the broad UV bump
itself does not allow us to uniquely identify
individual, specific molecules.
%
%

Finally, we note that PAHs are abundant and widespread
in the Universe, as revealed by a distinctive set of
emission bands at 3.3, 6.2, 7.7, 8.6, 11.3, and 12.7$\mum$
(see L\'eger \& Puget 1984, Allamandola et al.\ 1985,
Tielens 2008, Li 2020).
The 3.3$\mum$ PAH emission has recently been detected
by JWST's Mid-IR Instrument (MIRI) in SPT0418-47
at $z\approx 4.22$, a galaxy observed less than
1.5\,Gyr after the big bang (Spilker et al.\ 2023).
However, the origin of PAHs (and dust)
in JADES-GS-z6-0 is puzzling.
If star formation began when the Universe
was about 400\,Myr old, at the epoch of JADES-GS-z6-0,
even the oldest stars were only 400\,Myr old.
Therefore, there was not enough time
for low- to intermediate-mass stars
(about 0.5 to 8$\Msun$) to evolve sufficiently to
the dust-producing AGB phase.
In the Milky Way, AGB stars dominate the production
of stardust and, typically, it takes them
one billion years to evolve.
As only massive stars (with at least 8$\Msun$)
would have been able to evolve
in such a short timescale
(as that available for JADES-GS-z6-0),
it is often suggested that supernovae
were responsible for the dust
in the first billion years of cosmic time.
Nevertheless,  reverse shocks from supernovae
also destroy dust.
Apparently, the detection of the extinction bump
and the implied presence of PAHs in JADES-GS-z6-0
provide important insights into the stellar evolution
and dust condensation scenarios
in the earliest epochs of the Universe
(see Yang \& Li 2023).

\section{Summary}\label{sec:summary}
We reproduced the UV extinction bump
detected by JWST/NIRSpec at the cosmic dawn
in JADES-GS-z6-0 at redshift $z$\,$\approx$\,6.71,
with a mixture of compact, pericondensed PAH molecules
of different sizes. The UV bump seen in JADES-GS-z6-0
is narrower than any known Galactic or extragalactic
bump, and peaks at a wavelength considerably longer
than the nominal central wavelength of 2175$\Angstrom$.
While small graphite grains predict a UV bump
too broad to compare with that observed,
the weighted sum of the electronic absorption
spectra quantum chemically computed for
a number of PAH molecules closely explains
the UV bump of JADES-GS-z6-0,
although the UV bump alone does not allow one
to uniquely identify individual molecules.
We suggest that, in the early Universe, PAH molecules
originated from the physical and chemical processing
of carbonaceous (e.g., graphite) grains in the ISM,
which themselves originated from supernovae.

\begin{acknowledgements}
We thank B.T.~Draine, Q.~Li, C.E.~Mentzer,
Q.~Wang, and B.~Yang for stimulating discussions
and the anonymous referee for
helpful comments and suggestions.
QL and XJY are supported in part by
NSFC~12333005 and 12122302
and CMS-CSST-2021-A09.
JW gratefully acknowledges support from
the Cosmic Dawn Center through
the DAWN Fellowship.
The Cosmic Dawn Center (DAWN) is
funded by the Danish National
Research Foundation under grant No. 140.
\end{acknowledgements}


\makeatletter \renewcommand{\@biblabel}[1]{} \makeatother


\begin{thebibliography}{30}
\expandafter\ifx\csname natexlab\endcsname\relax\def\natexlab#1{#1}\fi

\bibitem[]{}Allamandola, L.J., Tielens, A.G.G.M., \& Barker, J.R.\
                 1985, ApJ, 290, L25

\bibitem[]{}Bohlin, R.~C., Savage, B.~D., \& Drake, J.~F.\
                 1978, ApJ, 224, 132

\bibitem[]{}Burkhardt, A.~M., Lee, K.L.K., Changala, P.B.,
                 et al.\ 2021, ApJL, 913, L18



\bibitem[]{}Cernicharo, J., Ag{\'u}ndez, M., Cabezas, C.,
                  et al.\ 2021, A\&A, 649, L15

\bibitem[]{}Clayton, G.C., Gordon, K.D., Salama, F.,
                  et al.\ 2003, ApJ, 592, 947

\bibitem[]{}Draine, B.T.\ 1993, in ASP Conf. Ser. 58,
                  The First Symposium on the Infrared Cirrus
                  and Diffuse Interstellar Clouds,
                  ed. R. Cutri \& W. Latter (San Francisco: ASP), 227

\bibitem[]{}Draine, B.T.\ 2003, ARA\&A, 41, 214


\bibitem[]{}Frisch, M. J., Trucks, G. W., Schlegel, H. B.,
                  et al\ 2016, Gaussian, Inc. 16, Revision C.01
                  (Wallingford CT: Gaussian, Inc.)


\bibitem[]{}Gredel, R., Carpentier, Y., Rouill{\'e}, G.,
                  Steglich, M., Huisken, F., \& Henning, Th.\
                  2011, A\&A, 530, A26

\bibitem[]{}Henning, Th., \& Schnaiter, M.\ 1998,
                  Earth, Moon \& Planets, 80, 179

\bibitem[]{}Iida, S., Hu, W., Zhang, R.,
                  et al.\ 2022, MNRAS, 514, 844

\bibitem[]{}J{\"a}ger, C., Mutschke, H., Henning, Th.,
                  \& Huisken, F.\ 2008, ApJ, 689, 249

\bibitem[]{}Joblin, C., L\'{e}ger, A., \& Martin, P.\ 1992,
                  ApJ, 393, L79

\bibitem[]{}Jones, A. P., Tielens, A. G. G. M.,
                  \& Hollenbach, D. J.\ 1996, ApJ, 469, 740

\bibitem[]{}Jones, R. O., \& Gunnarsson, O.\ 1989, RvMP, 61, 689

\bibitem[]{}Laporte, N., Ellis, R. S., Boone, F.\ 2017 ApJL, 837, L21

\bibitem[]{}L\'{e}ger, A., \& Puget, J.\ 1984, A\&A, 137, L5

\bibitem[]{}L\'eger, A., Boissel, \& d'Hendecourt, L.B.\
                 1988, Phys. Rev. Lett., 60, 921

\bibitem[]{}Li, A.\ 2020, Nature Astronomy, 4, 339

\bibitem[]{}Li, A., \& Draine, B.T.\ 2001, ApJ, 554, 778


\bibitem[]{}Li, Q., Yang, X. J., \& Li, A.\ 2024, MNRAS, 535, L58

\bibitem[]{}Lin, Q., Yang, X. J., \& Li, A.\ 2023, MNRAS, 525, 2380

\bibitem[]{}Malloci, G., Mulas, G., Cecchi-Pestellini, C.,
                  \& Joblin, C.\ 2008, A\&A, 489, 1183

\bibitem[]{}Marques, M.~A.~L., Castro, A., Bertsch, G.~F.,
                  et al.\ 2003, Computer Physics Communications, 151, 60

\bibitem[]{}Marques, M.~A.~L., \& Gross, E.~K.~U.\ 2004, ARPC, 55, 427

\bibitem[]{}McGuire, B.~A., Burkhardt, A.~M.,
                  Kalenskii, S., et al.\ 2018, Science, 359, 202

\bibitem[]{}McGuire, B.~A., Loomis, R.~A.,
                 Burkhardt, A.~M., et al.\ 2021,
                 Science, 371, 1265



\bibitem[]{}Pedrini, A., Adamo, A., Calzetti, D., et al.\
                  2024, ApJ, 971, 32


\bibitem[]{}Salama, F., Galazutdinov, G.~A., Kre{\l}owski, J.,
                  Biennier, L., Beletsky, Y., \& Song, I.-O.\ 2011,
                  ApJ, 728, 154

\bibitem[]{}Schnaiter, M., Mutschke, H., Dorschner, J.,
                  Henning, Th., \& Salama, F.\ 1998, ApJ, 498, 486

\bibitem[]{}Scott, A. D., Duley, W. W., \& Jahani, H. R.\
                  1997, ApJ, 490, L175

\bibitem[]{}Shivaei, I., Boogaard, L., D{\'\i}az-Santos, T.,
                  et al.\ 2022, MNRAS, 514, 1886

\bibitem[]{}Sita, M.~L., Changala, P.~B., Xue, C.,
                  et al.\ 2022, ApJL, 938, L12

\bibitem[]{}Spilker, J. S., Phadke, K. A., Aravena, M.,
                  et al.\ 2023, Nature, 618, 708

\bibitem[]{}Stecher, T. P.\ 1965, ApJ, 142, 1683

\bibitem[]{}Stecher, T. P., \& Donn, B.\ 1965, ApJ, 142, 1681

\bibitem[]{}Steglich, M., J{\"a}ger, C., Rouill{\'e}, G.,
                  Huisken, F., Mutschke, H., \& Henning, T.\
                  2010, ApJ, 712, L16

\bibitem[]{}Steglich, M., Carpentier, Y., J{\"a}ger, C.,
                  Huisken, F., R\"ader, H.-J., \& Henning, Th.\
                  2012, A\&A, 540, A110

\bibitem[]{}Stockett, M.~H., Bull, J.~N., Cederquist, H.,
                 et al.\ 2023, Nature Communications, 14, 395


\bibitem[]{}Tielens, A.G.G.M.\ 2008, ARA\&A, 46, 289

\bibitem[]{}Valencic, L.A., Clayton, G.C., \& Gordon, K.D.\
                  2004, ApJ, 616, 912


\bibitem[]{}Wang, Q., Yang, X. J., \& Li, A.\ 2023, MNRAS, 525, 983

\bibitem[]{}Wenzel, G., Cooke, I.~R., Changala, P.~B., et al.\ 2024a,
                Science, 386, 810

\bibitem[]{}Wenzel, G., Speak, T.~H., Changala, P.~B., et al.\ 2024b,
                Nature Astronomy, in press (arXiv:2410.00670)

\bibitem[]{}Witstok, J., Shivaei, I., Smit, R., et al.\
                 2023, Nature, 621, 267

\bibitem[]{}Yang, X.J., \& Li, A.\ 2023, Nature, 621, 260
%
\end{thebibliography}
\end{document}